\documentclass[a4paper]{article}

\usepackage{graphicx}
\usepackage{amssymb}
\usepackage{natbib}	

\title{Some considerations on skewness and kurtosis of vertical velocity in
the convective boundary layer}

\author{Alberto Maurizi and Francesco Tampieri\\
Institute of Atmospheric Sciences and Climate}

\begin{document}


\maketitle

\abstract{%
Data of skewness $S$ and kurtosis $K$ of vertical velocity in the convective
boundary layer from different datasets have been analysed. Vertical
profiles of $S$ were found to be grouped into two classes that display
different slopes with height: one is nearly constant and the other is
increasing. This behaviour can be explained using a simple model for the PDF
of vertical velocity and assuming two distinct vertical profiles of updraft
area fraction from literature. The possibility of describing the explicit
dependence of $K$ on $S$ was revised critically, also considering the
neutral limit as well as the limit for very small non-dimensional height. It
was found that the coefficients of the relationship depends on both the
Obukhov length scale $L$ and inversion height $z_i$.
}

\section{Introduction}

Although the empirical knowledge of atmospheric turbulence is increasing
steadily, a number of still unresolved questions remain, due in large part
to the intrinsic nature of the atmospheric boundary layer (ABL). In general,
a combination of field observations, laboratory and numerical experiments is
necessary to improve our understanding.

For instance, third- and fourth-order moments of vertical velocity are
difficult to measure because of the concurrent intrinsic unsteadiness of the
ABL on the one hand, and the need for long, steady samples for reliable
high-order
moment estimations on the other. Nevertheless, third-order moment data
have been considered at least since \citet{chiba-jmsj-1978} and, along with
fourth-order moments, continue to receive attention. Very reliable
data in neutral conditions have been collected for years in wind tunnel
experiments \citep[see][for instance]{durst_etal-tsf5-1987}. In the
convective boundary layer (CBL), a large amount of observations and large eddy simulations (LES)
are available: a summary can be found in \citet[][hereinafter
LLMSC]{lenschow_etal-blm-2012}.

Closures for Reynolds-averaged Navier-Stokes equation models of ABL are of
crucial importance in numerical weather prediction, as well as climate and atmospheric
composition modelling. As far the CBL is concerned, the need for closures
based on moments higher than 2 arises from the nonlocal nature of the
vertical mixing mechanism. A rather common approach is to solve equations for
the third-order moments, and use a closure for the fourth ones \citep[see,
for instance,][]{canuto_etal-jas-1994}.

The earliest closure on fourth-order moments was the quasi normal (QN)
approximation dating to \citet{millionshchikov-1941}. It assumes that
fourth-order cumulants of velocity correlations are zero as in the Gaussian
distribution, and expresses the principle of ``maximun
ignorance''\footnote{``... the one that makes the least claim to being
informed beyond the stated prior data, that is to say the one that admits
the most ignorance beyond the stated prior data''
\texttt{http://en.wikipedia.org/wiki/Principle\_of\_maximum\_entropy}}.
It was formulated for two point statistics (spatial correlation)
but can be simply applied to one point statistics, \textit{i.e.}, for
null separation. The QN approximation has been used for instance by
\citet{losch-grl-2004} and \citet{canuto_etal-om-2007}.

On purely statistical ground, the relationship $K \ge S^2 + 1$ defines the
region of existence of probability density functions (PDFs) in the $(S,K)$
space \citep[see, \textit{e.g.},][]{pearson-philtrans-1916,kendall-stuart}.
Consequently, the QN approximation is unrelizable for $S^2>2$. Based on this
consideration, \citet{maurizi-etc11-2007} suggested that the least biased
assumption on fourth-order statistics, a ``revised'' QN approximation, is
$K=3(S^2+1)$, which is always realizable and reduces to Gaussian for $S=0$.

Early attempts to organize third- and fourth-order moment data in the
$(S,K)$ space are found in \citet{durst_etal-tsf5-1987} for the neutral
boundary layer and in \citet{maurizi_etal-ae-1999} for different stability
conditions. A parabola-like relationship is often observed
in different turbulence regimes \citep{durst_etal-tsf5-1987,
shaw_etal-blm-1987, maurizi_etal-ae-1999} and is used to write closure
relationships \citep{alberghi_etal-jam-2002, gryanik_etal-jas-2002,
cheng_etal-jas-2005, gryanik_etal-jas-2005}. It is noteworthy that empirical
relationships and theoretical approaches (\textit{e.g.} based on mass-flux)
give basically the same results, consistent with the ``revised'' QN
hypothesis.

Moreover, the same qualitative results are found for data
of completely different nature \citep[see,
\textit{e.g.},][]{sattin_etal-plasmaphyscontrolfusion-2009,
cristelli_etal-pre-2012}, reflecting the purely statistical nature of the
relationship \citep{sattin_etal-physicascripta-2009}, which, due to the
curve that limits the existence of PDFs, produces a parabolic curvature of
the $(S,K)$ space. This also reflects on other properties,
\textit{e.g.}, Lagrangian correlation time and information entropy
\citep{maurizi_etal-ftc-2001,quan_etal-physicaa-2012}.

The purpose of this article is to extend the analysis by LLMSC of the closure
problem in $(S,K)$ space by reviewing the recent literature and bringing
together information from already available data. 
The different data are compared and merged in Section 2, where an
interpretation is given of the skewness behaviour for different convection
classes. Moreover, the behaviour very near the ground is analysed.
In Section 3, the skewness-kurtosis relationship is discussed in terms of
changing stability and distance from the wall (ground surface). In the final
Section, some conclusions are drawn.

\section{Bringing data together}

\subsection{Vertical profiles}
\begin{figure}
\includegraphics[width=0.5\textwidth]{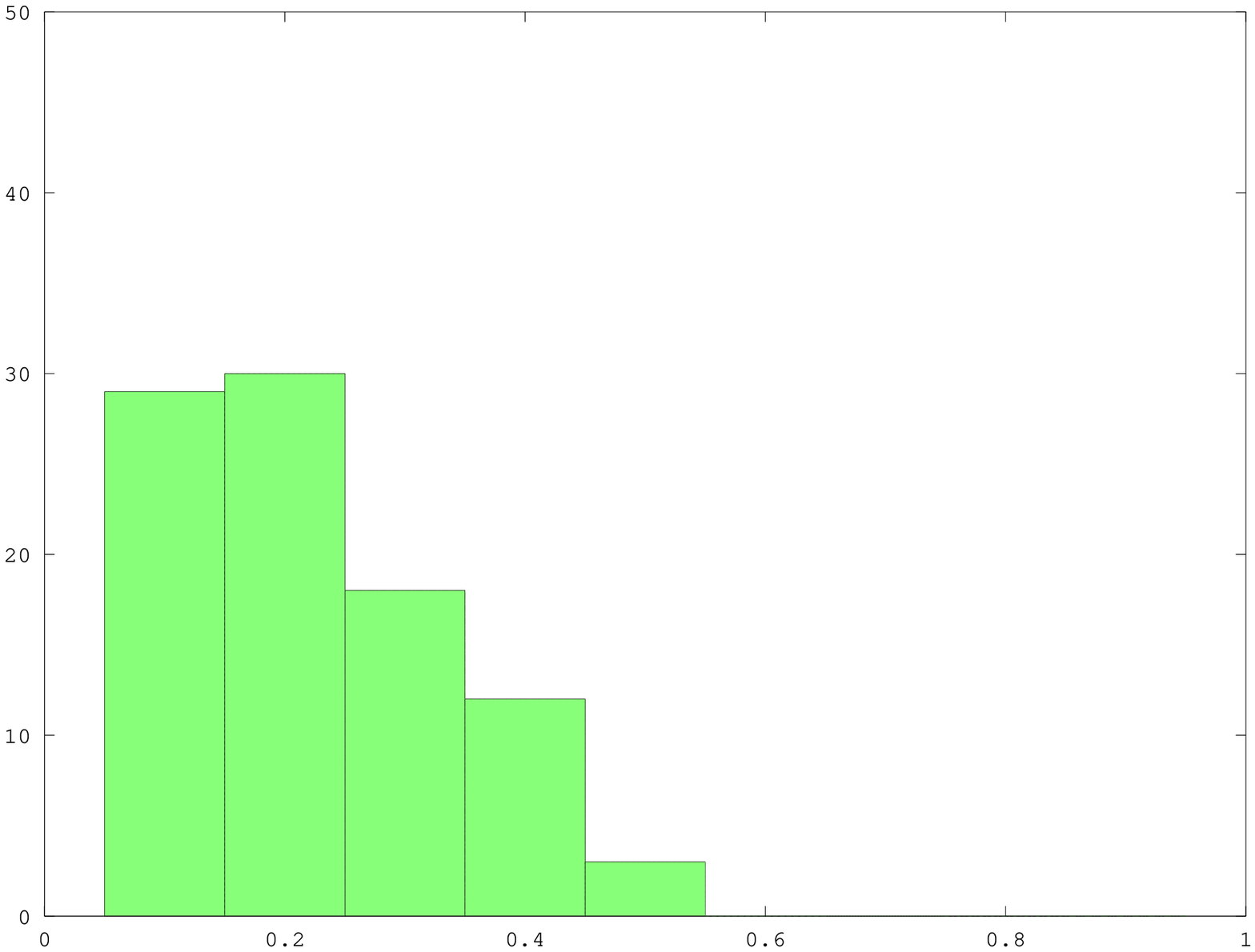}
\includegraphics[width=0.5\textwidth]{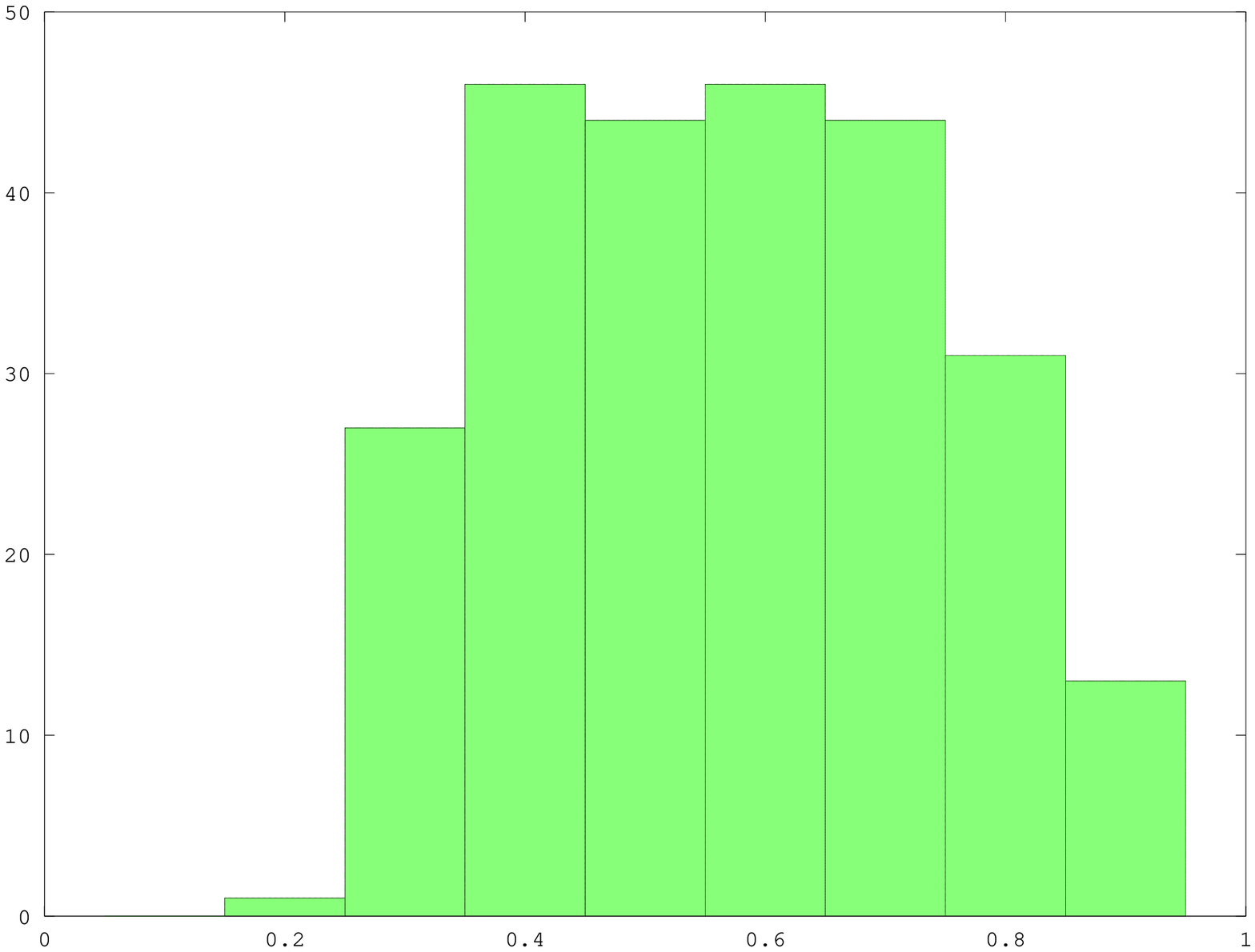}
\caption{Data distribution as a function of $z/z_i$: left TM, right LLMSC.}
\label{fig:distrib_height}
\end{figure}

LLMSC produced a very accurate vertical profile dataset of vertical
velocity in CBL measured by a LIDAR Doppler, also providing estimations of
stability parameters. Another dataset considered here is the set of
measurements taken in the Rome area \citep{mastrantonio_etal-blm-1994} by a
SODAR Doppler, which constitute the basis of the \citet{alberghi_etal-jam-2002}
and \citet{tampieri_etal-agit-2003} articles. The two datasets differ, in
particular, in the range of height covered. Figure~\ref{fig:distrib_height}
displays the number distribution of measurements with respect to the
normalised measurement height $z/z_i$ for the two datasets. The AMT dataset has
a 90\% quantile of 0.39, while the LLMSC one has a 10\% quantile of 0.34.
This highlights the complementarity of the two datasets.

\begin{figure}
\includegraphics{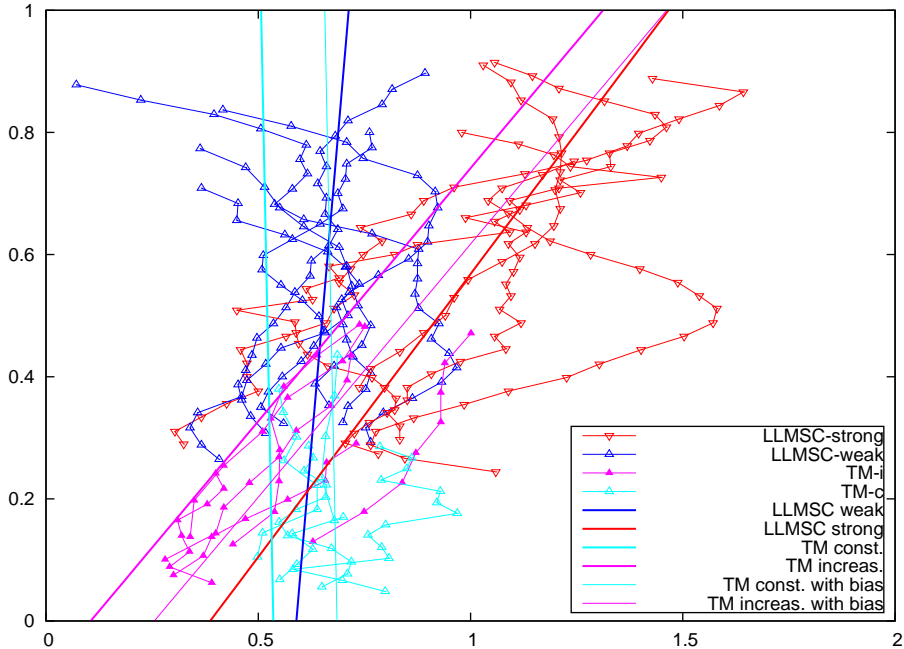}
\caption{$S(z/L)$ (left) and $K(z/L)$ (right) from LLMSC and TM. Fit are
performed on data with $z/z_i$ below 0.8. The second fit on TM data is
performed adding a bias of 0.15 to $S$.}
\label{fig:CBL_S_profiles}
\end{figure}

\begin{figure}
\includegraphics{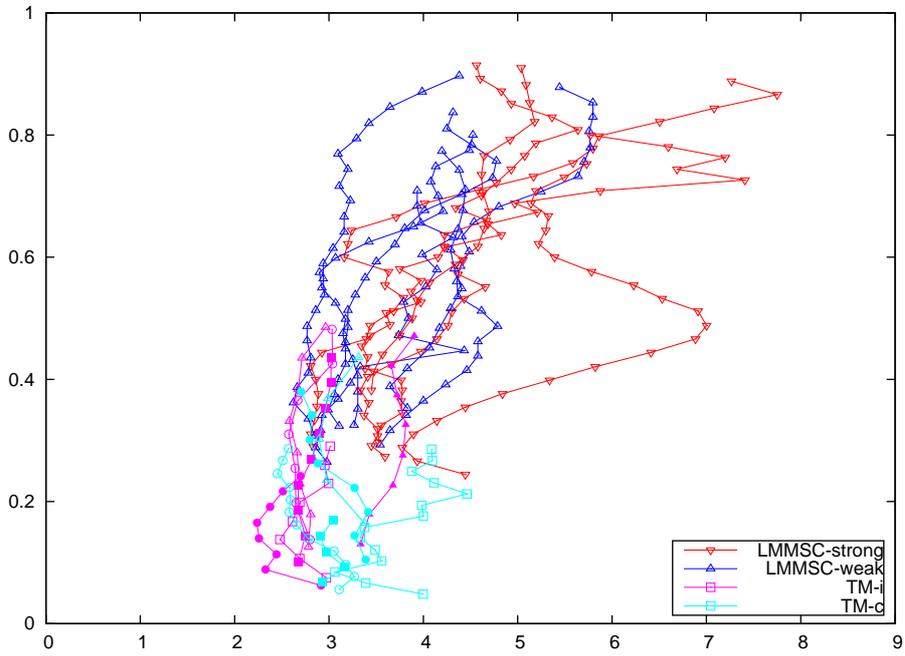}
\caption{As in Figure~\ref{fig:CBL_S_profiles} but for $K$ and without
fitting lines.}
\label{fig:CBL_K_profiles}
\end{figure}

Vertical profiles of $S$ and $K$ of both datasets are shown in
Figures~\ref{fig:CBL_S_profiles} and \ref{fig:CBL_K_profiles}. TM and LLMSC
provide independent classifications into two groups. The TM classification
is heuristically based on the trend of $S$ with $z/z_i$, for $z/z_i<0.5$.
The two classes are ``constant $S$'' and ``increasing $S$''. The LLMSC
classification is based on the convection strength (``weak'' and
``strong''), measured as $z_i/L$. Figure~\ref{fig:CBL_S_profiles} shows that
the two classifications are consistent. The ``constant $S$'' TM class
corresponds to the LLMSC ``weak convection'', while the ``increasing $S$''
class corresponds to LLMSC ``strong convection''. Independent linear fits of
the two datasets were performed for $z/z_i<0.8$. A similar slope was found
between LLMSC and TM for each class (see Figure~\ref{fig:CBL_S_profiles}).
The Figure clearly reveals a systematic difference ($\Delta S\simeq 0.15$)
between the two datasets, which could originate from the different measuring
systems: LIDAR and SODAR sample different volumes, leading to different cuts
in the turbulence spectrum.  It can also be observed that, according to TM,
the ``strong convection'' class displays a smaller $S$ for small $z/z_i$
than the ``weak convection'' class.

As far as $K$ is concerned, the two classes cannot be distinguished and the
two datasets are fully compatible.

A simple kinematic model can justify the two classes on the basis of the
vertical profile of the area fraction of updrafts $A$. Assuming the
\citet{baerentsen_etal-ae-1984} model for vertical velocity PDF, $S$ and $K$
can be expressed as a function of $A$ as 
\begin{equation}
S=\frac{2}{A(1-A)}(1-2A)
\label{eq:kinematic-S}
\end{equation}
and
\begin{equation}
K=\frac{5}{2}\frac{1-3A+3A^2}{A(1-A)}
\end{equation}
respectively.
The observed vertical profile of $A$ shows a certain degree of variability.
For instance, LLMSC suggest a model for the vertical velocity that implies a
constant $A$, while data from \citet{young-jas-1988} and the Large Eddy
Simulations of \citet{schumann_etal-jas-1991} support a weak variation with
height, with a minimum around $z/z_i=0.5$. According to
Equation~(\ref{eq:kinematic-S}), in the lower part of the CBL, decreasing $A$
corresponds to strongly increasing $S$, while $K$ displays smaller variations
with height. This behaviour is quite general and not related to the specific
PDF: it can be obtained even for the simplest case of two $\delta$
distributions. The kinematic model suggests that the vertical distribution
of the updraft area fraction is a function of $L/z_i$.

\subsection{Neutral limit}

\begin{figure}
\includegraphics{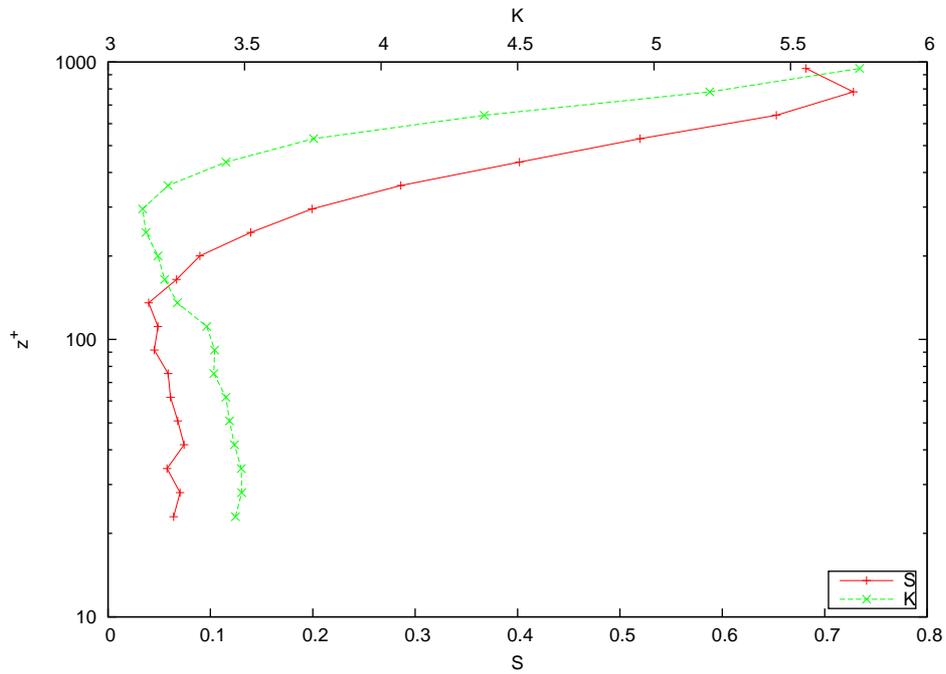}
\caption{Vertical profiles of $S$ and $K$ as a function of nondimensional
height $z^+=z/z_0$ in neutral conditions \citep{maurizi_etal-unpub-enflo}.}
\label{fig:enflo_SK_profiles}
\end{figure}

In order to investigate the neutral limit for skewness and kurtosis, new
data \citep{maurizi_etal-unpub-enflo} from a wind tunnel experiment
performed at
EnFlo\footnote{\texttt{http://www.surrey.ac.uk}},
are considered. Figure~\ref{fig:enflo_SK_profiles} shows that within the
surface layer, skewness of the vertical component of velocity in neutral
conditions is nearly constant at $S\simeq0.05$ and $K\simeq3.5$.

Data from the SGS2000 experiment using the ISFF facility\footnote{
\texttt{http://www.eol.ucar.edu/isf/facilities/isff/index.html}} analysed by
\citet{barberis-2007} add further complexity to the picture. Measurements
were performed over a relatively uniform surface, using two arrays of sonic
anemometers with height above the ground that was varied during the
experiment from 3.45 to 8.66 meters. A stationarity analysis was performed to
select periods long enough to make fourth-order statistics significant
\citep{barberis-2007}. These data cover a relatively large range of
stability, and a very small range of $z/z_i$ near the ground that cannot be
covered by remote sensing instruments such as LIDAR and SODAR.

\begin{figure}
\includegraphics{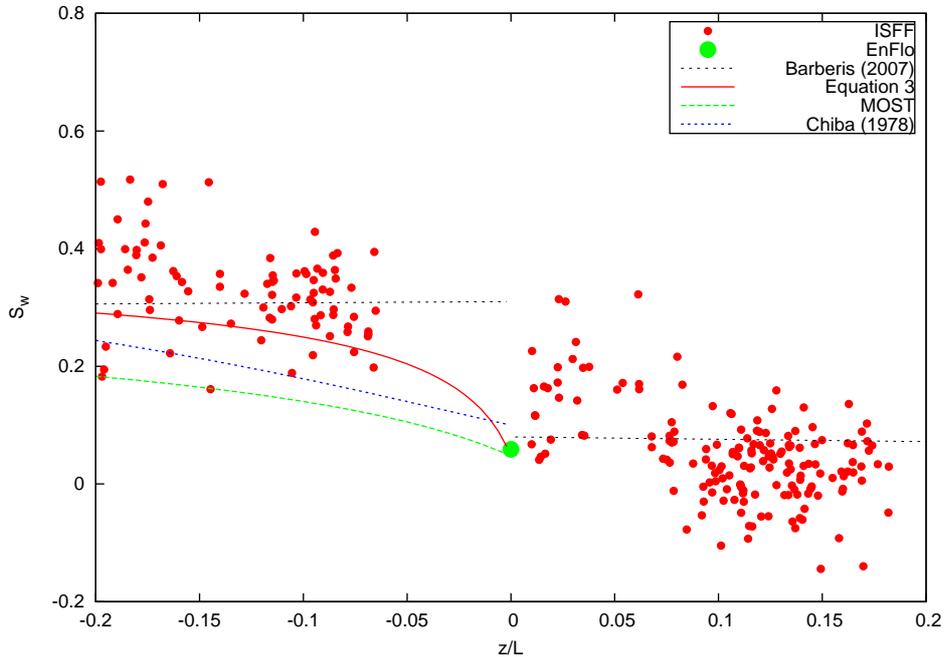}
\caption{Skewness of vertical velocity as a function of stability from ISFF
array \citep{barberis-2007}.}
\label{fig:isff}
\end{figure}

Because in the CBL $z/L \rightarrow 0$ as $z/z_i\rightarrow 0$, while
approaching the surface the flow tends to neutrality. Consequently, vertical velocity
statistics should approach the neutral $S$ and $K$ values. However, ISFF
data (Figure~\ref{fig:isff}) show that even in near neutral conditions a far
more pronounced asymmetry can appear for flows having small negative values
of $z/L$. The asymmetry introduced by a forcing mechanism like convection
far more marked than that caused by the presence of the wall.

The rather abrupt sharp transition from the convective value of $S$ to the
neutral one calls for a phenomenological analysis. Monin-Obukhov similarity
theory (MOST) suggests that vertical velocity variance $w^2$ scale on
$u_*^2$ for small $|z/L|$ and on $w_*^2 \propto u_*^2~|z/L|^{2/3}$ for free
convection conditions \citep{kader_etal-jfm-1990}. Analogous dependences can
be derived for $w^3$. However, many observations and theoretical
considerations suggest that the thickness of the boundary layer must enter
into
the scaling relationship, because of the influence of the largest eddies
even on the near surface values. As far as the horizontal components of
velocity are concerned, this point has been extensively discussed by
\citet{yaglom-pf-1994}. As a working hypothesis, it can be suggested that
a term depending on $z_i/L$ should also be included in the expression for the vertical
velocity variance.

An empirical expression for skewness based on the previous considerations is
\begin{equation}
S= \frac{c_{00} + c_0 |z_i/L| + c |z/L|}
{(b_{00} + b_0 |z_i/L|^{2/3} + b |z/L|^{2/3})^{3/2}}
\label{eq:s-yaglom}
\end{equation}
which reduces to standard MOST form setting $c_0$ and $b_0$ equal to 0.

From the literature, a possible choice of constants is: $b=3$, $b_{00}=1$,
$b_0=0.1$. Then, the third-order moment coefficients can be chosen in order
to match the asymptotic values of skewness for neutral conditions (0.05) and for
free convection (0.33).
In Figure~\ref{fig:isff}, Equation~(\ref{eq:s-yaglom}) is reported for the
case $c_0=0$ $b_0=0$ and for the complete expression with $z/z_i=500$. It
can be seen
that, in the second case, the drop is much sharper than in the
first one. Thus, the skewness behaviour can be qualitatively explained using
proper scaling relationships that account also for parameters not included
in MOST.  In Figure~\ref{fig:isff} the expression of
\citet{chiba-jmsj-1978} is also reported: it shows a smoother transition, similar
to our first case, consistent with the absence of the CBL height in the
formulation.

\section{Considerations on the S-K relationship}

\begin{figure}
\includegraphics{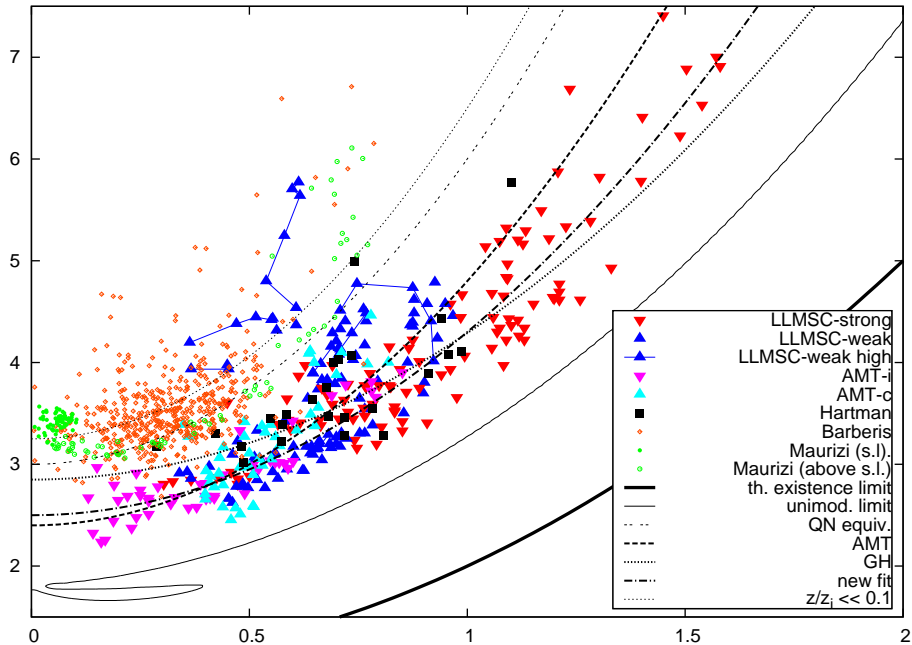}
\caption{Summary of $S$ and $K$ for different experiments. The LLMSC data
above $z/z_i=0.8$ are excluded. Curves are as described in the figure legend.
Data connected by lines are LLMSC weak convective cases for
$0.65<z/z_i<0.8$.}
\label{fig:SK_height}
\end{figure}

In $(S,K)$ space a lower bound to $K$ is known to exist: $K \geq S^2+1$ for
any PDF.  Less known is the fact that, due to the Gauss-Winckler inequality,
$K~\geq~1.8$ for a unimodal symmetric PDF \citep[][Vol I, p.
95]{kendall-stuart}. This result can be extended also to non-symmetric PDFs
($S\neq0$). Using the Gauss-Winckler inequality and substituing the
expression of non-centered moments in terms of centered ones, a relationship
between $K$ and a measure of the asymmetry can be established:
\begin{equation}
K\geq1.8+f(S,S_P)\,,
\label{eq:SK-unimodal}
\end{equation}
where $S_P$ is a measure of skewness, defined by Pearson as:
\begin{equation}
S_P=\frac{\mu_1 - m}{\sigma}
\end{equation}
in which $\mu_1$ is the mean of the distribution, $m$ is the (single) mode and
$\sigma$ is the standard deviation. Even if the explicit form of $f$ is not known in general, it can be
easily estimated for Pearson's family of (unimodal) PDFs as, in this
case, $S_P$ has a simple representation in terms of $S$ and $K$. This
particular form of $f$ is reported in Figure~\ref{fig:SK_height}. Although
this is not the most general one, it can be argued that different PDFs would
give a similar parabola-like relationship. In addition, numerical experiments
\citep{maurizi_etal-ftc-2001} confirm that PDFs based on simple models
become bi-modal while approaching the lower bound. Since two-value
velocity implies infinite accelerations, by extension, a highly bi-modal PDF
cannot be considered as a reliable representation of turbulence statistics.
Therefore, data are expected to lie well above the unimodal limit and models
should account for it.

The $S$ and $K$ data used so far are reported in Figure~\ref{fig:SK_height}.
AMT and LLMSC data cover different regions of $S$. This makes the fit
proposed by AMT, where $K\rightarrow2.4$ as $S\rightarrow0$, more suitable
for small $S$, while the \citet{gryanik_etal-jas-2002} fit, cited by LLSMC,
is more suitable for large $S$. Merging the two, a new fit can be proposed
in the form of
$K=a_2 S^2 + a_0$ with $a_2=1.8$ e $a_0=2.4$. The proposed relationship uses
two parameters as in \citet{gryanik_etal-jas-2002}, but provides a better
description of the small $S$ limit, according to AMT.  It is worth noting
that this parameterisation is appropriate only for data below
$z/z_i<0.65$. At higher levels it can be found (mainly for the weak
convective cases) that $K$ increases for decreasing (or constant) $S$, preventing
the possibility of describing $K$ as an explicit function of $S$ for the
whole CBL. Similar behaviour is observed also in LES (see LLMSC, their
Figure 11).

Another group of data (EnFlo and ISFF) clusters around a parabolic curve
with a different parameter: $a_0=3.3$. These data encompass both 
purely neutral and convective cases with $z/z_i\ll0.1$. It may be
argued that the coefficients of the parameterisations depend on both $z/L$
and $z/z_i$, with a dependence on $z/z_i$, strong for the 0 to 0.1 range
and very weak for $z/z_i>0.1$.

\section{Conclusions}

The observations regarding skewness and kurtosis in the convective boundary layer
analysed in this paper show that there is an intrinsic variability in
turbulence behaviour which cannot be simply explained in terms of
differences in surface fluxes.

The dependence of skewness on height is interpreted in terms of the area
distribution of updrafts and downdrafts, which is an important aspect of
convection physics \citep[an extensive discussion was given
by][]{hunt_etal-qjrms-1988}.
The correct representation of the vertical velocity PDF is considered a key
factor when dealing with dispersion problems, as well as in parameterisations
for numerical weather prediction models. Thus, the present results suggest
a careful consideration of these issues.

The transition from unstable to near neutral conditions in the surface layer
is sharper than expected from currently used similarity functions for
$\overline{w^2}$ and $\overline{w^3}$: the available data show values of
skewness around 3.3 in the surface layer for $z/L<-0.1$ and 0.05 for
$z/L=0$, with a substantial variability. The \citet{chiba-jmsj-1978} formula
did suggest a smoother transition. Also in this case, effects not
represented by the standard similarity approach, \textit{e.g.}, the
limitation to eddy size due to finite CBL height
\citep{kader_etal-jfm-1990}, may become important in shaping such
transitions.

Finally, the discussion about the $S$-$K$ relationship highlights that it
cannot be considered a universal one, but is likely to depend on the
structure and dynamics of the boundary layer under consideration.

The conclusions are that some tiles into the mosaic are added but also
questions for future research are raised.
  
\section*{Acknowledgements}
The software used for the production of this article (data analysis,
plotting, typesetting) is Free Software. The authors would like to thank the
whole free software community, the Free Software Fundation
(\texttt{http://www.fsfs.org}) and the Debian Project
(\texttt{http://www.debian.org}).

\bibliographystyle{personal}
\bibliography{bibtex.bib}

\begin{thebibliography}{31}
\expandafter\ifx\csname natexlab\endcsname\relax\def\natexlab#1{#1}\fi

\bibitem[Alberghi et~al., 2002]{alberghi_etal-jam-2002}
Alberghi, S., A.~Maurizi, and F.~Tampieri, 2002: Relationship between the
  vertical velocity skewness and kurtosis observed during sea-breeze
  convection. \textit{J. Appl. Meteorol.}, \textbf{41}, 885--889.

\bibitem[Baerentsen and Berkowicz, 1984]{baerentsen_etal-ae-1984}
Baerentsen, J.~H. and R.~Berkowicz, 1984: Monte-carlo simulation of plume
  diffusion in the convective boundary layer. \textit{Atmos. Environ.},
  \textbf{18}, 701--712.

\bibitem[Barberis, 2007]{barberis-2007}
Barberis, E., 2007: Analisi statistiche nello strato limite turbolento, thesis,
  Univ. Torino, Dip. Fisica.

\bibitem[Canuto et~al., 2007]{canuto_etal-om-2007}
Canuto, V.~M., Y.~Cheng, and A.~Howard, 2007: Non-local ocean mixing model and
  a new plume model for deep convection. \textit{Ocean Modelling}, \textbf{16},
  28--46.

\bibitem[Canuto et~al., 1994]{canuto_etal-jas-1994}
Canuto, V.~M., F.~Minotti, C.~Ronchi, and R.~Ypma, 1994: Secon-order closure
  pbl model with new third-order moments: comparison with les data.
  \textit{Journal of the Atmospheric Sciences}, \textbf{51}, 1605--1618.

\bibitem[Cheng et~al., 2005]{cheng_etal-jas-2005}
Cheng, Y., V.~M. Canuto, and A.~Howard, 2005: Nonlocal convective pbl model
  based on new third- and fourth-order moments. \textit{Journal of the
  Atmospheric Sciences}, \textbf{62}, 2189--2204.

\bibitem[Chiba, 1978]{chiba-jmsj-1978}
Chiba, O., 1978: Stability dependence of the vertical velocity skewness in the
  atmospheric surface layer. \textit{Journal of the meteorological society in
  Japan}, \textbf{56}, 140--142.

\bibitem[Cristelli et~al., 2012]{cristelli_etal-pre-2012}
Cristelli, M., A.~Zaccaria, and L.~Pietronero, 2012: Universal relation between
  skewness and kurtosis in complex dynamics. \textit{Physical Review E},
  \textbf{85}.

\bibitem[Durst et~al., 1987]{durst_etal-tsf5-1987}
Durst, F., J.~Jovanovic, and L.~J. Kanevce, 1987: Probability density
  distribution in turbulent wall boundary-layer flows, \textit{Turbulent Shear
  Flows 5}, F.~Durst, B.~E. Launder, J.~L. Lumley, F.~W. Schmidt, and J.~H.
  Whitelaw, eds., Springer Verlag, Berlin.

\bibitem[Gryanik and Hartmann, 2002]{gryanik_etal-jas-2002}
Gryanik, V. and J.~Hartmann, 2002: A turbulence closure for the convective
  boundary layer based on a two-scale mass-flux approach. \textit{Journal of
  the Atmospheric Sciences}, \textbf{59}, 2729--2744.

\bibitem[Gryanik et~al., 2005]{gryanik_etal-jas-2005}
Gryanik, V., J.~Hartmann, S.~Raasch, and M.~Schroter, 2005: A refinement of the
  millionshchikov quasi-normality hypothesis for convective boundary layer
  turbulence. \textit{Journal of the Atmospheric Sciences}, \textbf{62},
  2632--2638.

\bibitem[Hunt et~al., 1988]{hunt_etal-qjrms-1988}
Hunt, J. C.~R., J.~C. Kaimal, and J.~E. Gaynor, 1988: Eddy structure in the
  convective boundary layer - new measurements and new concepts. \textit{Quart.
  J. Roy. Meteor. Soc.}, \textbf{114}, 827--858.

\bibitem[Kader and Yaglom, 1990]{kader_etal-jfm-1990}
Kader, B.~A. and A.~M. Yaglom, 1990: Mean fields and fluctuation moments in
  unstably stratified turbulent boundary layers. \textit{J. Fluid Mech.},
  \textbf{212}, 637--662.

\bibitem[Kendall and Stuart, 1977]{kendall-stuart}
Kendall, S.~M. and A.~Stuart, 1977: \textit{The Advanced Theory of Statistics},
  vol.~1, 4th ed., C. Griffin \& Co., London.

\bibitem[Lenschow et~al., 2012]{lenschow_etal-blm-2012}
Lenschow, D.~H., M.~Lothon, S.~D. Mayor, P.~P. Sullivan, and G.~Canut, 2012: A
  comparison of higher-order vertical velocity moments in the convective
  boundary layer from lidar with in situ measurements and large-eddy
  simulation. \textit{Boundary-Layer Meteorol}, \textbf{143}, 107–123.

\bibitem[Losch, 2004]{losch-grl-2004}
Losch, M., 2004: On the validity of the millionshchikov quasi-normality
  hypothesis for open-ocean deep convection. \textit{Geophysical Research
  Letters}, \textbf{31}.

\bibitem[Mastrantonio et~al., 1994]{mastrantonio_etal-blm-1994}
Mastrantonio, G., A.~Viola, S.~Argentini, G.~Fiocco, L.~Giannini, L.~Rossini,
  G.~Abbate, R.~Ocone, and M.~Casonato, 1994: Observations of sea-breeze events
  in rome and the surrounding area by a network of doppler sodars.
  \textit{Boundary Layer Meteorology}, \textbf{71}, 67--80.

\bibitem[Maurizi, 2007]{maurizi-etc11-2007}
Maurizi, A., 2007: Quasi normal hypothesis revised, \textit{Advances in
  Turbulence XI,Proceedings of the 11th EUROMECH European Turbulence
  Conference}, vol. 117 of \textit{Springer Proceedings in Physics}, Springer,
  pp. 603--605.

\bibitem[Maurizi and Lorenzani, 2001]{maurizi_etal-ftc-2001}
Maurizi, A. and S.~Lorenzani, 2001: Lagrangian time scales in inhomogeneous
  non-{Gaussian} turbulence. \textit{Flow, Turbulence and Combustion},
  \textbf{67}, 205--216.

\bibitem[Maurizi and Robins, 2000]{maurizi_etal-unpub-enflo}
Maurizi, A. and A.~Robins, 2000: Boundary-layer flow and dispersion over a
  two-dimensional hill; high-order statistics of the flow and concentration
  fields, experiment performed at EnFlo, UniSurrey, UK.

\bibitem[Maurizi and Tampieri, 1999]{maurizi_etal-ae-1999}
Maurizi, A. and F.~Tampieri, 1999: Velocity probability density functions in
  {Lagrangian} dispersion models for inhomogeneous turbulence. \textit{Atmos.
  Environ.}, \textbf{33}, 281--289.

\bibitem[Millionshchikov, 1941]{millionshchikov-1941}
Millionshchikov, M.~D., 1941: Theory of homogeneous isotropic turbulence.
  \textit{Dokl. Akad. Nauk SSSR}, \textbf{32}, 611--614.

\bibitem[Pearson, 1916]{pearson-philtrans-1916}
Pearson, K., 1916: Mathematical contributions to the theory of evolution. xix:
  second supplement to a memoir on skew variation. \textit{Phil.Trans.Roy. Soc.
  A}, \textbf{216}, 432.

\bibitem[Quan et~al., 2012]{quan_etal-physicaa-2012}
Quan, L., E.~Ferrero, and F.~Hu, 2012: Relating statistical moments and entropy
  in the stable boundary layer. \textit{Physica A}, \textbf{391}, 231–247.

\bibitem[Sattin et~al., 2009{\natexlab{a}}]{sattin_etal-physicascripta-2009}
Sattin, F., m.~agostini, r.~cavazzana, g.~serianni, p.~scarin, and n.~vianello,
  2009{\natexlab{a}}: About the parabolic relation existing between the
  skewness and the kurtosis in time series of experimental data.
  \textit{PHYSICA SCRIPTA}, \textbf{79}, 045006.

\bibitem[Sattin et~al.,
  2009{\natexlab{b}}]{sattin_etal-plasmaphyscontrolfusion-2009}
Sattin, F., m.~agostini, p.~scarin, n.~vianello, r.~cavazzana, L.~Marrelli,
  g.~serianni, S.~J. Zweben, R.~J. Maqueda, Y.~Yagi, H.~Sakakita, H.~Koguchi,
  S.~Kiyama, Y.~Hirano, and J.~L. Terry, 2009{\natexlab{b}}: On the statistics
  of edge fluctuations: comparative study between various fusion devices.
  \textit{Plasma Phys. Control. Fusion}, \textbf{51}, 055013.

\bibitem[Schumann and Moeng, 1991]{schumann_etal-jas-1991}
Schumann, U. and C.-H. Moeng, 1991: Plume fluxes in clear and cloudy convective
  boundary layers. \textit{Journal of Atmospheric Sciences}, \textbf{48},
  1746--1757.

\bibitem[Shaw and Seginer, 1987]{shaw_etal-blm-1987}
Shaw, R.~H. and I.~Seginer, 1987: Calculation of velocity skewness in real and
  artificial plant canopies. \textit{Boundary-Layer Meteorology}, \textbf{39},
  315--332.

\bibitem[Tampieri and Maurizi, 2003]{tampieri_etal-agit-2003}
Tampieri, F. and A.~Maurizi, 2003: Investigations on convective boundary layer
  turbulence using sodar data. \textit{Annals of Geophysics}, \textbf{46},
  451--457.

\bibitem[Yaglom, 1994]{yaglom-pf-1994}
Yaglom, A.~M., 1994: Fluctuation spectra and variances in convective turbulent
  boundary layers: a reevaluation of old models. \textit{Phys. of Fluids},
  \textbf{6}, 962--972.

\bibitem[Young, 1988]{young-jas-1988}
Young, G.~S., 1988: Turbulence structure of the convective boundary layer. part
  ii: Phoenix 78 aircraft observations of thermals and their environment.
  \textit{Journal of Atmospheric Sciences}, \textbf{45}, 727--735.

\end{thebibliography}

\end{document}